\documentclass[12pt]{article}
\usepackage{amsmath, amsthm}

\newtheoremstyle{theorem}
  {10pt}		  
  {10pt}  
  {\sl}  
  {\parindent}     
  {\bf}  
  {. }    
  { }    
  {}     
\theoremstyle{theorem}
\newtheorem{theorem}{Theorem}

\newtheoremstyle{defi}
  {10pt}		  
  {10pt}  
  {\rm}  
  {\parindent}     
  {\bf}  
  {. }    
  { }    
  {}     
\theoremstyle{defi}
\newtheorem{definition}{Definition}

\begin{document}

\title{Logarithmic Cycles of the Time: An Quaternionic Approach.}
\author{J. Mar\~ao\\
$^1$ State University of Maranh\~ao - UEMA\\
              Mathematic and Informatic Department - DEMATI\\
              S\~ao Lu\'is - MA\\
              650000-001\\
              and\\
              Federal University of Maranh\~ao - UFMA,\\ B.Sc. Interdisciplinary Science and Technology - (BCT).\\
65085-580 - Maranh\~ao - BRAZIL\\
josemarao@cecen.uema.br\\[2pt]}

\maketitle

\begin{abstract}
The satisfactory development of Quaternionic Analysis has indicated new solutions for physical and mathematical problems. It is worth mentioning the fact that quaternions possess four dimensions, and in this way they may be considered as natural elements for the formulation many of physical and geometrical problems. The scope of this paper is to show that the cycles of time \cite{Penrose} may be performed quaternionic logarithm.
\end{abstract}

\section{Introduction}
\label{intro}
The time analysis has been grounded throughout the years \cite{Penrose} shows the possibility of a cyclical cosmological time. However, he did not show a standard mathematical foundation for his arguments, neither, early, outlined possible future developments for this new view of the universe.\\
Thus, this work is intended to formulate in terms of quaternionic logarithms \cite {LogTrig} some of the ideas discussed in \cite{Penrose}. In this context some important results were already previously addressed in the work \cite{HyperCycles}, through exponential functions. However, the aim of this work is to show the possibility of approaching the cosmological cycles of time, to logarithmic functions and present  that the end, whether or not there is an advantage in this approach, taking by quaternionic logarithms instead of exponential hypercomplex function to expand a new view of the universe.\\
The approach here will be based on submission of quaternionic exponential and logarithmic functions, and then use the latter approach in the proposed problem on this paper. It is essential to emphasize that the space where the study is analyzed in the quaternionic space $H$ and the functions are functions of quaternionic variables taking values $q\in H.$ Thus, throughout the text the quaternionic exponential function and logarithmic functions will be called simply quaternionic exponential function and logarithmic function

\section{The Exponential and Logarithmic Quaternionic Functions.}
Seeking for a better support to this work, a definition is essential for the subsequent understanding. Then follows the definition of quaternionic function.
\begin{definition}A quaternionic function is a law $f: H \longrightarrow H$ that associates to each $q=(q_{1},q_{2},q_{3},q_{4})$ one $w=f(q)$ in the division algebra of quaternions, is represented as follows: \\
\begin{eqnarray}
f(q_{1},q_{2},q_{3},q_{4}) = f_{1}(q_{1},q_{2},q_{3},q_{4}) + f_{2}(q_{1},q_{2},q_{3},q_{4})i\\
+ f_{3}(q_{1},q_{2},q_{3},q_{4})j + f_{4}(q_{1},q_{2},q_{3},q_{4})k.\nonumber
\end{eqnarray}
\end{definition}
Considering $q\in H$ an quaternionic number given $q=q_{1}+q_{2}i+q_{3}j+q_{4}k,$ or $q=q_{1}+\vec{q},$ as shown in \cite{LogTrig}. The exponential function is given by:
\begin{equation}e^{q}=e^{q_{1}}\{cos|\vec{q}|+\vec{q}(\frac{sin|\vec{q}|}{|\vec{q}|})\}.\end{equation}
The expression above allows a number of conclusions concerning the mentioned function; one of them is that it is Hyper Periodic, and this fact is illustrated geometrically by means of a cube which edges measure $2\pi.$ The fact that the exponential function belongs to this region makes it possible to nominate the region and considering that the region is critical to the quaternionic exponential function region.\\
Another important function in the development of the Quaternionic Analysis, the quaternionic logarithmic function is implemented \cite{LogTrig}. Then, it is first presented the generalized system of spherical coordinates in four dimensional, as bellow \cite{LogTrig}:
$$u_{1}=rcos\theta_{1}cos\theta_{2}cos\theta_{3}, 0<r<\infty$$
$$u_{2}=rcos\theta_{1}cos\theta_{2}sin\theta_{3},0<\theta_{3}<2\pi$$
$$u_{3}=rcos\theta_{1}sin\theta_{2}, 0<\theta_{2}<\frac{\pi}{2}$$
$$u_{4}=rsin\theta_{1}, 0<\theta_{1}<\frac{\pi}{2}$$
Identifying, in the expression (1):
$$e^{q}=e^{q_{1}}\{cos|\vec{q}|+\vec{q}(\frac{sin|\vec{q}|}{|\vec{q}|})\}$$
the value of $e^{\vec{q}},$ is given by:
\begin{equation}\phi=(cos|\vec{q}|,q_{2}\frac{sin|\vec{q}|}{|\vec{q}|},q_{3}\frac{sin|\vec{q}|}{|\vec{q}|},q_{4}\frac{sin|\vec{q}|}{|\vec{q}|})\end{equation}
or
\begin{equation}\phi=cos|\vec{q}|+q_{2}\frac{sin|\vec{q}|}{|\vec{q}|}i+q_{3}\frac{sin|\vec{q}|}{|\vec{q}|}j+q_{4}\frac{sin|\vec{q}|}{|\vec{q}|}k,\end{equation}
where replacing $(2)$ by $(1)$ there is:
\begin{equation}e^{q}=e^{u_{1}}\phi.\end{equation}
Identifying the $(4)$ expression in spherical coordinates we have:
$$e^{q}=e^{u_{1}}(cos\theta_{1}cos\theta_{2}cos\theta_{3}+cos\theta_{1}cos\theta_{2}sin\theta_{3}i+cos\theta_{1}sin\theta_{2}j+sin\theta_{1}k)$$
is logarithm of $q$ will be denoted by $lnq$ and will be the inverse of the exponential function. Thus $ w=lnq,$ satisfies the relation:
$$e^{w}=q,$$
where $q\neq 0.$ Now let $w\in H$ indicated by $w=w_{1}+w_{2}i+w_{3}j+w_{4}k$ or $w=(w_{1},w_{2},w_{3},w_{4}k)$ and using the generalized spherical coordinates:
$$u'_{1}=rcos\theta_{1}cos\theta_{2}cos\theta_{3}, 0<r<\infty$$
$$u'_{2}=rcos\theta_{1}cos\theta_{2}sin\theta_{3},0<\theta_{3}<2\pi$$
$$u'_{3}=rcos\theta_{1}sin\theta_{2}, 0<\theta_{2}<\frac{\pi}{2}$$
$$u'_{4}=rsin\theta_{1}, 0<\theta_{1}<\frac{\pi}{2}.$$
Now $w$ is given by:
$$w=u'_{1}+u'_{2}i+u'_{3}j+u'_{4}k$$
$$w=r(\frac{u'_{1}}{r}+\frac{u'_{2}}{r}i+\frac{u'_{3}}{r}j+\frac{u'_{4}}{r}k)$$
$$w=r(cos\theta_{1}cos\theta_{2}cos\theta_{3}+cos\theta_{1}cos\theta_{2}sin\theta_{3}i+cos\theta_{1}sin\theta_{2}j+sin\theta_{1}k)$$
where the value $r$ is positive, i.e, $r>0.$ Thus,
$$e^{w}=e^{u'_{1}}\{cos|\vec{u'}|+\vec{u'}(\frac{sin|\vec{u'}|}{|\vec{u'}|})\}$$
where $w=u'_{1}+\vec{u'}.$\\
However,
$$e^{u'_{1}}e^{\vec{u'}}=q$$
taking $e^{\vec{u'_{1}}}=r,$
\begin{equation}u'_{1}=ln|r|\end{equation}
and
$$\vec{u'}=ln(cos\theta_{1}cos\theta_{2}cos\theta_{3}+cos\theta_{1}cos\theta_{2}sin\theta_{3}i+cos\theta_{1}sin\theta_{2}j+sin\theta_{1}k)$$
but $w=u_{1}+\vec{u'}.$\\
Therefore, there is an expression for the logarithmic quaternionic function.
\begin{equation}w=lnq=ln|r|+ln(cos\theta_{1}cos\theta_{2}cos\theta_{3}+cos\theta_{1}cos\theta_{2}sin\theta_{3}i+cos\theta_{1}sin\theta_{2}j+sin\theta_{1}k)\end{equation}
\section{Logarithmic Cycles of the Time.}
The implementation appeared in the previous section suggests that the logarithm is used as a way to show the cyclicity of some physical quantity. At the present time it will show that is cyclicity happens simply considering that $q=t+xi+yj+zk,$ $q\in H$ is a function $\eta(q)=\eta(t,x,y,z),$ given by:
\begin{equation}\eta(t+\tau i,x,y,z)=(t+\tau i)+xi+yj+zk\end{equation}
or
\begin{equation}\eta(t+\tau i,x,y,z)=(t+\tau i)+\vec{u}.\end{equation}
Thus, the above function is characterized by its position, given by the vector $\vec{u}=(x,y,z)$ and time, here denoted by $t.$ The function can determine the temporal evolution of some physical quantity or simply modeling the approach to the problem.\\
Now, the function $\eta(q),$ in logarithmic form, can be written on using $(5),$ $u'_{1}=t+\tau i=ln|r|,$ and $(6),$ as follows:
\begin{equation}ln[\eta(q)]=ln|e^{t+\tau i}|+ln(cos\theta_{1}cos\theta_{2}cos\theta_{3}+cos\theta_{1}cos\theta_{2}sin\theta_{3}i+cos\theta_{1}sin\theta_{2}j+sin\theta_{1}k),\end{equation}
or,
\begin{equation}ln[\eta(q)]=ln|e^{t}(cos(\tau)+ isin(\tau)|+ln(cos\theta_{1}cos\theta_{2}cos\theta_{3}+cos\theta_{1}cos\theta_{2}sin\theta_{3}i+cos\theta_{1}sin\theta_{2}j+sin\theta_{1}k),\end{equation}
Therefore, considering the quaternionic function of the form: \begin{equation}\eta[q_{\varphi}]=[t+(2k\pi\tau)+(\theta_{1}+2k\pi)i+(\theta_{2}+2k\pi)j+(\theta_{3}+2k\pi)k],\end{equation}
it becomes clear that:
\begin{equation}ln[\eta(q)]=ln[\eta(q_{\varphi})].\end{equation}
The above results can be summarized in the following theorem:
\begin{theorem}If $q, q_{\varphi}\in H$ are quaternionic numbers of the form $q=(t,x,y,z)$ and $q_{\varphi}=(t+\tau i,x,y,z),$ furthermore $ln(\eta(q))$ is the logarithmic quaternionic function, then
\begin{equation}ln[\eta(q)]=ln[\eta(q_{\varphi})].\end{equation}
\end{theorem}

\section{Conclusion.}
The \textbf{Theorem} \textbf{1} shows the possibility of making the cyclicality in the variable $\tau$ and the spatial variables, which are located here by $\theta_{1},$ $\theta_{2}$ and $\theta_{3},$ and it is possible to resume the initial function. The following statements are concerned with what was presented here:
\begin{enumerate}
\item If the function $\eta$ representing a physical phenomenon, the cyclicality of time allows, since the spatial dimensions occur simultaneously, unlike what happens in the case considered in \cite{HyperCycles} where only time was considered cyclical.
\item The regions of variation of $\tau$ and the spatial variables represented here by generalized angles in spherical coordinates can be interpreted as the fundamental range $-\pi<\tau\leq\pi$ and the cube edges parallel to the coordinate axes and length $2\pi$ centered at the origin of the three dimensional system.
\item The possibility of cyclicity in logarithmic form allows two disjoint approaches: one to separate the space and time to another time and space in cyclic variations.
\end{enumerate}
Therefore, the model presented here can be applied to dependent physical models of time and space coordinates and it was shown that so there Cycles of the Time is necessary that the time variable is of the form $t+\tau i$ which can be considered a flat composition of times so that a one is dependent on of another second the complex relationship presented, and more, these remain after one cycle in exponential models \cite{HyperCycles} and logarithmic unchanged and return to the starting position. Furthermore, considering the probable connected model variables studied simultaneously and a cycle time will result in a cycle of space, both because they are in sync formulation present in log cycle time coupled to cycles of space.
\section{Acknowledgments}
My family.

\end{document}